\documentclass[aps,prl,twocolumn,showpacs]{revtex4}
\usepackage{epsfig}

\usepackage{amsmath,amssymb}
\newcommand{\ignore}[1]{}

\begin{document}

\title{Systems with two symmetric absorbing states: relating the 
microscopic dynamics with the macroscopic behavior}

\author{Federico Vazquez}
\email[E-mail: ]{federico@ifisc.uib-csic.es}
\author{Crist\'obal L\'opez}
%\email[E-mail: ]{clopez@ifisc.uib.es}

\affiliation{IFISC, Instituto de F\'{\i}sica Interdisciplinar y Sistemas
Complejos (CSIC-UIB), E-07122 Palma de Mallorca, Spain}
\homepage{http://www.ifisc.uib.es}

\date{\today}

\begin{abstract} We propose a general approach to study spin models with two
symmetric absorbing states.  Starting from the microscopic dynamics on a
square lattice, we derive a Langevin equation for  the time evolution of the
magnetization field, that successfully explains coarsening properties of a
wide range of  nonlinear voter models and systems with intermediate states.
We find that the  macroscopic behavior only depends on the first derivatives
of the spin-flip probabilities.  Moreover, an  analysis of the  mean-field
term reveals the three types of transitions commonly observed in these systems
-generalized voter, Ising and directed percolation-.  Monte Carlo simulations
of the spin dynamics qualitatively agree with theoretical predictions.
\end{abstract}


\maketitle

\emph{Introduction.}  The non-equilibrium dynamics of interacting particle
systems with absorbing states is a central issue in modern Statistical
Mechanics \cite{Hinrichsen00,Odor03}.  In the last years much interest has
been given to the special case of models with two symmetric ($Z_2$-symmetry)
absorbing states~\cite{Hammal05,Droz03,Dornic01}, due to the relevance in many
different contexts of a subclass of them, the so-called voter models.  They
have been widely used in diverse disciplines to study different dynamics, such
as species competition~\cite{Clifford73}, allele frequency in
genetics~\cite{Baxter07}, kinetics of heterogeneous
catalysis~\cite{Krapivsky92-Frachebourg96}, and more recently, opinion
formation~\cite{Castellano07} and language  spreading~\cite{Abrams03}.  In
these models, the state of a particle in a lattice site evolves according  to
the density of states in its near neighborhood. Interesting dynamical
behaviors arise depending on the specific updating rules, the number of states
and the functional form of the transition probabilities between
configurations.  For instance, it has been found that the addition of memory
or inertia in the spin  dynamics \cite{Dall'Asta07,Stark08}, the introduction
of intermediate states  \cite{Castello06,Baronchelli06,Dall'Asta08}, or the
use of non-linear transitions \cite{Schweizer}, result in a drastic  change of
the coarsening properties and final outcome of the system.

Despite that the dynamical rules of the models are  very
different in nature, many of them seem to share the same macroscopic
behavior, such as coarsening and criticality.  However, the minimal conditions 
that a microscopic dynamics must hold in order to observe a particular 
behavior have not been clearly identified yet.  In other words, given a spin 
model defined by its flipping transition probability and interaction range, 
can we anticipate how the system will evolve over time ? 

In this article, we try to answer this question by developing an approach that
connects the microscopic dynamics with the macroscopic space-time evolution of
the system in square lattices.  We derive a Langevin equation for the
magnetization field, and find that, at the macroscopic level, the properties
are only determined by the first three derivatives of the transition 
probabilities.

Our Langevin equation coincides with that postulated by Hammal et. al.
\cite{Hammal05} by symmetry arguments, but now the coefficients of the
different terms have a clear explanation in terms of the  transition
probabilities.  The analysis of this equation helps to understand some of the
open questions about phase ordering in these systems, that is, whether the
coarsening is driven by curvature like in the Ising model \cite{Bray02}, or it
is without surface tension like in the original voter model (VM) 
\cite{Dornic01}.
This approach also explains, from a different perspective than
in~\cite{Dall'Asta08}, why adding intermediate states to the VM leads
to an effective surface tension.  Moreover, numerical simulations of the
spin dynamics with different interaction ranges confirm the three
possible classes of phase transitions unveiled in~\cite{Hammal05}, and
clarifies an apparent controversy found between previous works
\cite{Dornic01,Droz03}.

\emph{The model and the Langevin equation.}  Each 
site ${\bf r}= (r_1,..,r_d)$
of a $d$-dimensional square lattice is occupied by one particle with a spin
that can assume either value  $1$ (up) or $-1$ (down).  The dynamics consists
of choosing, at each time step, a site ${\bf r}$ at random and flipping the
spin $S_{\bf r}$ at this site with a probability  that is a function, $f(-
S_{\bf r} \psi_{\bf r})$, of the product between $S_{\bf r}$ and the
particle's local magnetization  $\psi_{{\bf r}} \equiv \frac{1}{z} \sum_{{\bf
r'/r}}S_{{\bf r'}}$, where the sum is over the $z$ nearest-neighboring sites
${\bf r'}$ of ${\bf r}$.  $z$ is an arbitrary integer number that defines the
interaction range (e.g. $z=4$ for first-nearest neighbor interactions). In
order to ensure that the fully ordered configurations $S_{\bf r} = 1$ or
$S_{\bf r}=-1$ for all {\bf r} are absorbing, the \emph{flipping probability}
$f$ must vanish when the spin is aligned with all its neighboring  spins, i.e,
$f(-1)=0$.

We want to derive a Langevin equation for the field  $\phi_{\bf r}(t)$, that 
is a continuous representation of the spin at site  ${\bf r}$, at
time $t$.   For this we follow a standard approach (see 
\cite{Dall'Asta08}), and consider an  ensemble of  $\Omega$ copies of the
system, each copy representing a particular spin configuration.  This is
equivalent to assume $\Omega$ spin particles at each site of the lattice (our
microscopic model corresponds exactly to $\Omega=1$, but this substitution can
be made at the end of the calculation).  In this approach
$\phi_{\bf r}(t)$ is replaced by the average spin value  $\phi_{\bf r}(t) \to
\frac{1}{\Omega} \sum_{j=1}^\Omega S_{\bf  r}^j$, and $\psi_{\bf r}$ by the
average local field  $\psi_{\bf r} \to \frac{1}{z} \sum_{{\bf r'/r}}\phi_{\bf
r'}(t)$, at site  ${\bf r}$ at time $t$.  In a time step, a site ${\bf x}$ and
one particle  from that site with spin $S_{\bf x}$ are randomly chosen.   Then
the spin attempts to flip with probability $f(-S_{\bf x} \psi_{\bf x})$.  If
the flipping occurs, then the field on the  entire lattice, represented as
$\{\phi\}$, changes only at site  ${\bf x}$ by $-2 S_{\bf x}/\Omega$.  Thus
the rising and lowering  transition  rates are $W\left(\{\phi\} \to~\{\phi\}
\pm \frac{2}{\Omega} \delta_{{\bf x},{\bf r}}\right)= W^{\pm}(\phi,{\bf x},t)
=  \frac{1}{2} \left( 1\mp  \phi_{\bf x} \right) f(\pm \psi_{\bf x})$.   We
can write a master equation for the time evolution of the probability
distribution $\mathcal P(\{\phi\},t)$, which after an expansion to second
order in $1/\Omega$ leads to the following  Fokker-Planck equation
\begin{eqnarray}
&&\frac{\partial}{\partial t} \mathcal P(\{\phi\},t)= \\ 
&& \sum_{\bf r} -\frac{1}{\Omega} \frac{\partial}{\partial \phi } 
\Bigl\{ 2 \left[ W^+(\phi,{\bf r},t)- W^-(\phi,{\bf r},t) \right] 
\mathcal P(\{\phi\},t) \Bigr\}
\nonumber \\
&& +\frac{1}{\Omega^2} \frac{\partial^2}{\partial \phi^2} \Bigl\{2 
\left[W^+(\phi,{\bf r},t)+W^-(\phi,{\bf r},t) \right] \mathcal P(\{\phi\},t) 
\Bigr\}. \nonumber
\end{eqnarray}
From the above equation, using the expressions for  $W^{\pm}(\phi,{\bf r},t)$
and rescaling time with $\Omega$, we arrive to the Langevin equation
\begin{equation}
\frac{\partial \phi_{\bf r}(t)}{\partial t} = \left[1-\phi_{\bf r}(t)\right] 
f(\psi_{\bf r})-
\left[1+\phi_{\bf r}(t)\right] f(-\psi_{\bf r}) + \eta_{\bf r}(t),
\label{dphi-dt}
\end{equation}
where $\eta_{\bf r}(t)$ is a Gaussian white noise with correlations
\begin{eqnarray}
&& \langle \eta_{\bf r}(t) \eta_{\bf r'}(t') \rangle = 
 \Bigl\{ \left[1-\phi_{\bf r}(t)\right]  
f(\psi_{\bf r}) \nonumber \\ 
&&+ \left[1+\phi_{\bf r}(t)\right] f(-\psi_{\bf r}) \Bigr\} 
\delta_{\bf r,r'} \delta(t-t')/{\Omega^{1/2}}. 
\end{eqnarray}
Note that up to now our derivation is completely general, and the fact that
the system has two absorbing states is only present in the  condition for the
flipping probability.  Now we look for an approximation to  Eq.(\ref{dphi-dt})
which, however, captures the behavior of a wide range of absorbing $Z_2$
models.  As  explained in \cite{Hammal05}, this is obtained with a $\phi^6$
model, i.e. when the right hand side of Eq.~(\ref{dphi-dt}) is proportional to
$\phi^5$. Thus, we expand $f$ around $\psi_{\bf r}=0$ up to forth order in
$\psi_{\bf r}$, but also making $f$ vanish at $\psi_{\bf r} = -1$ (the
condition for existence of absorbing states at $\pm 1$):
\begin{equation}
\label{f}
f(\psi_{\bf r}) = \frac{1}{2} (1+\psi_{\bf r}) \left(c+a \psi_{\bf r}+
d \psi_{\bf r}^2- b \psi_{\bf r}^3 \right),
\end{equation}
where the real coefficients $a, b, c$ and $d$ are, for convenience, defined as
(primes denoting derivatives)
\begin{eqnarray}
\label{coef}
c \equiv 2 f(0), ~ a \equiv 2 f'(0)- c,~ \nonumber \\ 
d \equiv f''(0)-a,~ b \equiv -\frac{f'''(0)}{3}+d .  
\end{eqnarray}
To obtain a closed equation for $\phi_{\bf r}(t)$, we replace  
expression~(\ref{f}) for $f(\psi_{\bf k})$ into
Eq.~(\ref{dphi-dt}) and make the substitution $\psi_{\bf r} = \phi_{\bf r} +
\Delta \phi_{\bf r}$, where we define the Laplacian operator 
$\Delta \phi_{\bf r} \equiv  \frac{1}{z} \sum_{\bf r'/r} \left( \phi_{\bf
  r'}-\phi_{\bf r} \right) =\psi_{\bf r}-\phi_{\bf r}$.  We then expand 
to first order in $\Delta \phi_{\bf r}$
(assuming that $\phi_{\bf r}$ is a smooth  function of ${\bf r}$ in the long  
time
limit, so that $\Delta \phi_{\bf r} \ll \phi_{\bf r} < 1$), and obtain the 
following Langevin equation for $\phi_{\bf r}$ 
\begin{eqnarray}
\label{dphi-dt1}
\frac{\partial \phi}{\partial t}&=&(1-\phi^2) (a \phi - b \phi^3) \nonumber \\ 
&+&\left[a+c+(d-2a-3b)\phi^2 \right] \Delta \phi + \eta, 
\end{eqnarray}
with correlations for the noise 
\begin{eqnarray}
\label{corr}
&&\langle \eta_{\bf r}(t) \eta_{\bf r'}(t') \rangle = \\  
&&\Bigl\{ (1-\phi^2) (c+d\phi^2)+(a-c+2d)\phi \Delta \phi \Bigr\} 
\delta_{\bf r,r'} \delta(t-t'), \nonumber 
\end{eqnarray}
where $\phi$ denotes $\phi_{\bf r}(t)$.
Equations~(\ref{dphi-dt1}) and (\ref{corr}) agree  with the Langevin
equation proposed in \cite{Hammal05}, based on symmetry arguments, to
describe order-disorder phase transitions in general models  with two
symmetric absorbing states (the noise correlation in their equation
is a simplified version of Eq.(\ref{corr})).  We have derived this expression
from  the microscopic dynamics, and therefore the different parameters of the
theory have a clear interpretation as a function of the transition rates.
The first two terms of Eq.~(\ref{dphi-dt1}) can also be obtained by
identifying the field  $\phi_{\bf r}$ with the average value $\langle S_{\bf
r} \rangle$ of the spin  at site ${\bf r}$ over all spin configurations, and
following the moments  approach technique used by Krapivsky et al. for the
original VM \cite{Krapivsky92-Frachebourg96}.  This corresponds to
the  $\Omega \to \infty$ limit, for which fluctuations are neglected and the
equation for $\phi$ becomes deterministic.

\begin{figure}[t] 
\begin{center}
\includegraphics[width=3.4in, clip=true]{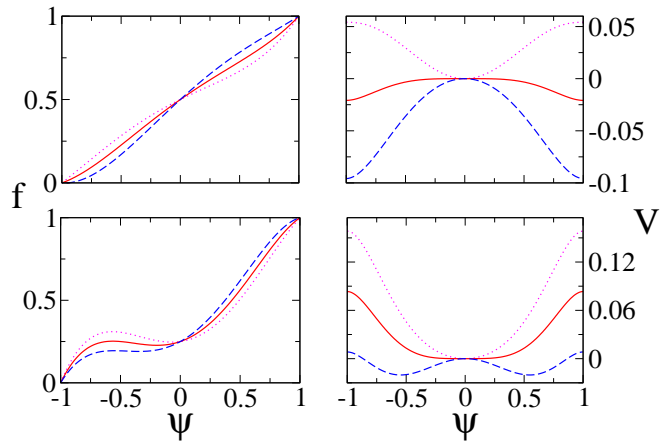}
\end{center}
\caption{Flipping probability $f$ (left boxes) and its associated potential $V$
  (right boxes) vs local field $\psi$. Curves correspond to coefficient values 
  $a=-0.3$ (dotted), $a=0$ (solid), $a=0.3$ (dashed), for $b=-0.25$ (top) and
   $b=1.0$ (bottom).  For both values of $b$, a single-well and double-well
  potentials are obtained for $a<0$ and $a>0$ respectively.}
\label{f-phi}
\end{figure}

We now use Eq.~(\ref{dphi-dt1}) to gain insight into the macroscopic
ordering dynamics.  At the mean-field (MF) level, where the noise term is
neglected,  Eq.~(\ref{dphi-dt1}) takes the form of a time-dependent
Ginzburg-Landau equation \cite{Bray02} $\frac{\partial \phi}{\partial t} =  D
\Delta \phi -\frac{\partial V}{\partial \phi}$, with  the potential
$V(\phi)=-\frac{a}{2} \phi^2 + \frac{a+b}{4} \phi^4 - \frac{b}{6} \phi^6$ and
$D$ an effective diffusion constant.  In Fig.~\ref{f-phi} we sketch the shape
of $V$ and its associated $f(\psi)$ for different values of $a$ and $b$.   
When $a>0$ ($f'(0)>f(0)$), $V$ has two symmetric  minima, thus the
system coarsens driven by surface tension \cite{Bray02}.   On the contrary,
when $a<0$, ($f'(0)<f(0)$), the minimum is at $\phi=0$, then the system 
remains in an active disordered state with  particles continuously flipping 
their spins, and a global magnetization that  fluctuates around zero.

To illustrate our previous results, we now analyze a general class of 3-state
models \cite{Castello06,Baronchelli06}, known to exhibit curvature 
driven  by
surface tension, as recently shown in \cite{Dall'Asta08}.  They are
composed  by two external absorbing states $S=\pm 1$, and an intermediate
state $S=0$.  The transition of a particle from  $S=-1$ to $S=1$ happens in
two stages.  If we denote by  $\rho_-, \rho_0$ and $\rho_+$, the densities of
nearest-neighboring particles  in states $-1,0$ and $1$, respectively, the
particle first switches from $S=\mp 1$ to $S=0$ with probability $\rho_{\pm} +
\rho_0/2$, and then with the same probability $\rho_{\pm} + \rho_0/2$ from 
$S=0$
to $S=\pm 1$.  Hence, disregarding the intermediate state, this model can be
thought as one with two absorbing states $S=\pm 1$, and an effective
transition  probability from $S=-1$ to $S=1$ equals to $(\rho_+ +
\rho_0/2)^2$.  In the  same  way, we believe that models with many
intermediate states behave as  equivalent 2-state models with effective
transition probabilities that are  \emph{ non-linear} in the local densities,
and our theory can be applied.  We test this by considering a 2-state model
proposed by Abrams and Strogatz to study the competition between two
languages~\cite{Abrams03}. The transition probabilities are given by  $P(\mp 1
\to \pm 1) = \rho_{\pm}^q$, where $q$ is a positive real number, so that the
$q=1$ case reduces to the original VM, while $q \ge 2$ corresponds to models
with one or more intermediate states.  In terms of the local field  $\psi = 2
\rho_+-1$,  the flipping probability  is  $f(\psi) =
\left(\frac{1+\psi}{2}\right)^q$.  Calculating the coefficients  defined in
Eq.~(\ref{coef}), and replacing them  into Eq.~(\ref{dphi-dt1}) we obtain
\begin{eqnarray}
\frac{\partial \phi}{\partial t} &=& \frac{(q-1)}{3 \times 2^{q}}(1-\phi^2) 
\left[ 6 \phi + (q-2)(q-3) \phi^3 \right] \nonumber \\ 
&+& \frac{q}{2^{q}}
\left[ 2+ (q-1)(q-4) \phi^2 \right] \Delta \phi + \eta.
\end{eqnarray}
When $q<1$ ($f'(0)<f(0)$), the stable solution is $\phi=0$, corresponding to a
disordered active system. When $q>1$ ($f'(0)>f(0)$), the stable solutions are
$\phi=\pm 1$, thus the system orders driven by surface tension until it
reaches one of the absorbing states ($\phi=1,-1$ for all ${\bf r}$).   In
particular, this type of ordering is observed when $q=2$, for which the
Langevin equation has a similar form as the one derived by Dall'Asta et al.
for 3-state models
\begin{equation}
\frac{\partial \phi}{\partial t} = \frac{1}{2} \left(\phi - \phi^3 \right) +
(1-\phi^2) \Delta \phi + \sqrt{1-\phi^2}~\eta,
\end{equation} 
confirming the above mentioned equivalence with a 2-state model with quadratic
transition probability.  The special case $q=1$ corresponds to the VM, with 
Langevin equation
\begin{equation}
\frac{\partial \phi}{\partial t} = \Delta \phi +  \sqrt{1-\phi^2-\Delta \phi}
~\eta.
\end{equation} 
Neglecting the Laplacian in the noise term, this equation is the same as the
one suggested by Dickman et al. \cite{Dickman95}.  Even though the potential
is zero, the system still orders due to the presence of the Laplacian, but
without surface tension.

\emph{Numerical simulations of the microscopic dynamics.}
Equation~(\ref{dphi-dt1}) shows that, at the MF level, the macroscopic
behavior of a particular model defined by the probability $f$ is only
determined by the coefficients $a$ and $b$, that are expressed in terms of the
derivatives of $f$ around $\psi_{\bf k}=0$.  As qualitatively predicted by the
MF theory, and confirmed in \cite{Hammal05}, for a fix value $b<b^*$, there is
a unique generalized voter (GV) order-disorder transition at a critical value  
$a_{GV}$, that
separates an active stationary state with absolute  magnetization $m=0$ for
$a<a_{GV}$, from a frozen ordered state with $m=1$ for $a>a_{GV}$.  For
$b>b^*$, as $a$ is increased, a symmetry breaking Ising transition is observed
at a value $a_I$, followed by a directed percolation (DP) transition at a
value $a_{DP}>a_I$.  For $a<a_I$ the system is disordered ($m=0$), whereas for
$a_I<a<a_{DP}$ it gets partially magnetized ($0<m<1$).  Above $a_{DP}$, the
system relaxes to the fully ordered state ($m=1$).  Since we  know the
connection between the macroscopic and the microscopic dynamics  expressed in
Eq.~(\ref{dphi-dt1}) and Eq.~(\ref{f}) respectively, we now study  these
transitions by a Monte  Carlo simulation of the model. This approach is
complementary to the one followed by Hammal \emph{et al.} in which they
integrate the Langevin equation; and it also allows to test the field theory.

We performed numerical simulations on a $2$-dimensional square lattice with
first-nearest neighbors (1st-NNs) interactions ($z=4$).  We used flipping
probabilities that are polynomial functions of the form of Eq.~(\ref{f}), for
$b=-0.25$, $0.5$ and $3.0$, and various values of $a$ (see Fig.~\ref{f-phi}).
Coefficients $c$  and $d$ were set in order to make $f$ an increasing function
of $\psi$, and to  arbitrarily fix the point  $f(\psi=1)=1$.   Starting from
an ordered configuration of down spins (initial quenching), we flipped the
spins of four neighboring sites at the center of the lattice and let the
system evolve.  We found that the average density of up spins $N$ and the
survival probability $P$, for the three values of $b$, decay at the
critical transition point $a_{GV}$ as  $N \sim t^{\eta}$ and 
$P \sim t^{-\delta}$ (not shown), 
where   $\eta \simeq 0$ and $\delta \simeq 0.95$ agree with the
exponents $0$  and $1.0$ respectively, of the GV universality
class~\cite{Odor03,Dickman95}.  Also, we found that the average density of
interfaces has, at $a_{GV}$, a logarithmic decay with time ($\rho \simeq
\pi/(2 \ln t)$), as in the VM \cite{Krapivsky92-Frachebourg96}.
Surprisely, only a  GV transition was observed for all three values of $b$,
in contradiction with \cite{Hammal05}.

\begin{figure}[t]
\begin{center}
\includegraphics[width=3.2in, clip=true]{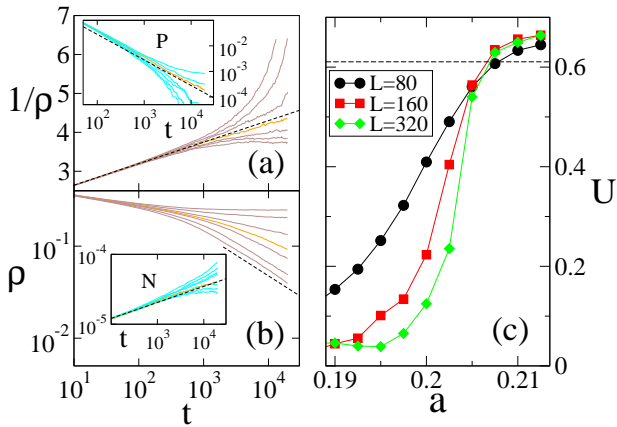}
\end{center}
\caption{GV, DP and Ising transitions on a $2D$ square lattice of side $L=400$
with up to 3rd-NNs ($z=12$).  (a) $1/\rho$ vs time on a log-linear scale, for
$b=-0.25$  and values of $a$ around the GV critical point $a_{GV} \simeq
-0.1105$. (b)   $\rho$ vs time on a log-log scale for $b=0.5$ and values of
$a$ around the DP critical point $a_{DP} \simeq 0.2127$.   (c) Binder cumulant
$U$ vs $a$ for $b=0.5$. Curves  cross at $a_I \simeq 0.205$, where $U \simeq
0.56$, close to  the universal value $0.6107$ of the $d=2$ Ising model
(horizontal dashed line). Insets: (a) survival probability $P$ and (b) density
of up spins $N$, on a $800^2$ lattice,  starting from a quenched
configuration.  Curves are  averages over $10^5$ realizations. Dashed straight
lines have slopes $2/\pi$ and $-1$ in (a) and its inset respectively, whereas
the slopes are $-0.45$ and $0.2295$ in (b).}
\label{12NN}
\end{figure}  

However, our results are in agreement with that of Dornic et. al. 
\cite{Dornic01}
where, by studying the dynamics of coarsening without surface tension (also
with 1st-NNs interactions), they conjectured that all models with
$Z_2$-symmetry and without bulk noise exhibit GV transitions.  This apparent
disagreement between theory (three transitions) and simulations (only GV
transition) comes from the fact that Ising
dynamics is observed when there is bulk noise, and this happens in our model
when the interaction distance is equal or larger than $2$.  Thus, a spin
surrounded by $8$ parallel spins can  still flip if at least one of the four
3-rd NNs is antiparallel.  Indeed,  simulations taking up to 2-nd NNs ($z=8$)
also revealed GV only, but  increasing the interactions up to 3-rd NNs
($z=12$), all GV, Ising and DP transitions were observed.  This agrees with
the work by Droz et al. \cite{Droz03}, in which they studied an absorbing
Ising model in two dimensions, and found that extending the interaction range
up to $z=12$ neighbors, the transition from disorder to order splits into a
first Ising transition  that breaks the symmetry, and then a DP transition to
the unique absorbing  state selected by the spontaneous symmetry breaking.

In Fig.~\ref{12NN} we plot the numerical results for $z=12$ neighbors.  We see 
that for $b=-0.25$ (Fig.\ref{12NN}(a)), the decay of $\rho$ and $P$
at $a_{GV} \simeq -0.1105$ correspond to that of a GV transition, whereas for
$b=0.5$ (Fig.~\ref{12NN}(b)), the transition to complete order happens at a 
value $a_{DP} \simeq
0.2127$ at which  $\rho \sim t^{-\delta}$, $N \sim t^{\eta}$ and $P \sim
t^{-\delta}$ (not  shown), with $\delta \simeq 0.45$ and $\eta \simeq 0.2295$,
i.e. DP critical exponents.  In order to find the Ising transition, we
calculated the Binder cumulants $U=1-m_4/3 m_2^2$ (Fig.~\ref{12NN}(c)), where
$m_4$ and $m_2$ are the fourth and second moments of the magnetization, as a
function of $a$.  As we can see, at the critical point
$a_I \simeq 0.205$, the curves of the Binder cumulants for different system
sizes cross each other at the value $U \simeq 0.56$, similar to the universal
value $0.6107$ of the $2D$ Ising model.

\emph{Conclusions.}  Summing up, we have derived from the microscopic
dynamics, the Langevin equation for the magnetization field of general
non-equilibrium spin systems with two symmetric absorbing states.  This
equation agrees with the one introduced in previous work, but now the
dependence of the different terms on the flipping probability is explicitly
stated.   This methodology allows to predict the macroscopic behavior, such as
critical properties and ordering dynamics, by simply knowing the derivatives
of the transition probabilities.   A large class of models in many different
disciplines can be studied in this way.  The generalization of this approach
to models with an arbitrary number of symmetric absorbing states seems to be
challenging.

We are very grateful to Maxi San Miguel and Miguel A. Mu\~noz for fruitful
discussions.  We acknowledge  support from project FISICOS (FIS2007-60327) of
MEC and  FEDER, and NEST-Complexity project PATRES (043268).

%%%%%%%%%%%%%%%%%%%%%%%%%%%%%%%%%%%%%%%%%%%%%%%%%%%%%%%%%%%%%%%%%%%%%%%%%%%%%%%

{}

%%%%%%%%%%%%%%%%%%%%%%%%%%%%%%%%%%%%%%%%%%%%%%%%%%%%%%%%%%%%%%%%%%%%%%%%%%%%%%%


\begin{thebibliography}{}

\bibitem{Hinrichsen00} H. Hinrichsen, Adv. Phys. {\bf 49}, 815 (2000).

\bibitem{Odor03} G. \'Odor, Rev. Mod. Phys. {\bf 76}, 663 (2004).   

\bibitem{Hammal05} O. Al Hammal, H. Chat\'e, I. Dornic, and M. A. Mu\~noz, 
  Phys. Rev. Lett. {\bf 94}, 230601 (2005).

\bibitem{Dornic01} I. Dornic, H. Chat\'e, J. Chave, and H. Hinrichsen,
  Phys. Rev. Lett. {\bf 87}, 045701 (2001).

\bibitem{Droz03} M. Droz, A. L. Ferreira, and A. Lipowski, Phys. Rev. E 
  {\bf 67}, 056108 (2003).

\bibitem{Clifford73} P. Clifford and A. Sudbury, \emph{Biometrika}, 
  {\bf 60}, 581 (1973). 

\bibitem{Baxter07} G. J. Baxter, R. A. Blythe and A. J. MacKane, Math. Biosci. 
  {\bf 209}, 124 (2007).

\bibitem{Krapivsky92-Frachebourg96} P. L. Krapivsky, Phys. Rev. A {\bf 45}, 
  1067 (1992); \\L. Frachebourg and P. L. Krapivsky, Phys. Rev. E {\bf 53}, 
  R3009 (1996).

\bibitem{Castellano07} C. Castellano, S. Fortunato and V. Loreto, 
{\tt arXiv:0710.3256} (2007). 

\bibitem{Abrams03} D. M. Abrams and S. H. Strogatz, Nature {\bf 424}, 900 
(2003).

\bibitem{Dall'Asta07} L. Dall'Asta and C. Castellano, Europhys. Lett. 
{\bf  77}, 60005 (2007).

\bibitem{Stark08} H.-U. Stark, C. J. Tessone, and F. Schweitzer,
  Phys. Rev. Lett. {\bf 101}, 018701 (2008).  

\bibitem{Castello06} X. Castell\'o, V. M. Egu\'iluz and M. San Miguel, 
  New Journal of Physics {\bf 8}, 308 (2006).

\bibitem{Baronchelli06} A. Baronchelli, L. Dall'Asta, A. Barrat, and
  V. Loreto, Phys. Rev. E {\bf 73}, R015102 (2006).

\bibitem{Dall'Asta08} L. Dall'Asta and T. Galla, {\tt arXiv:0806.0817} (2008).

\bibitem{Schweizer} F. Schweitzer and L. Behera, {\tt arXiv:cond-mat/0307742}
  (2008).

\bibitem{Bray02} A. J. Bray, Adv. Phys. {\bf 51}, 481 (2002).

\bibitem{Dickman95} R. Dickman, and A. Yu. Tretyakov, Phys. Rev. E {\bf 52},
  3218 (1995).

\end{thebibliography}
\end{document}